\documentclass[twocolumn,aps,prd,showpacs,superscriptaddress,nofootinbib,amsmath,amssymb,floats,floatfix,showkeys,notitlepage,longbibliography]{revtex4-1}

\usepackage{comment}
\usepackage{graphicx}
\usepackage[]{mdframed}
\usepackage{subfigure}
\usepackage{palatino}
\usepackage[commandnameprefix=always]{changes}
\usepackage{hyperref}
\hypersetup{colorlinks=true,linkcolor=blue,urlcolor=blue,citecolor=blue}
\usepackage[toc,page]{appendix}
\usepackage[normalem]{ulem}

\usepackage{orcidlink}
\usepackage{lipsum}
\usepackage{graphicx}
\usepackage{subfigure}
\usepackage{palatino}
\usepackage{sans}
\usepackage{adjustbox}
\usepackage{latexsym}
\usepackage{amsmath}
\usepackage{amssymb}
\usepackage{amsfonts}
\usepackage{dcolumn}
\usepackage{bm}
\usepackage{tikz}
\usepackage{bigints}
\usepackage{array,tabularx,multirow,booktabs}
\usepackage[tracking=true]{microtype}
\SetTracking{}{500}
\SetTracking{encoding={*}, shape=sc}{40}
\UseRawInputEncoding 
\allowdisplaybreaks
\usepackage{adjustbox}
\usepackage{latexsym}
\usepackage{amsmath}
\usepackage{amssymb}
\usepackage{amsfonts}
\usepackage{dcolumn}
\usepackage{bm}
\usepackage{tikz}
\usepackage{bigints}
\usepackage{array,tabularx,multirow,booktabs}
\usepackage[tracking=true]{microtype}
\usepackage{color}
\UseRawInputEncoding 
\allowdisplaybreaks

\begin{document} \sloppy

\title{General Approach on Shadow Radius and Photon Spheres in Asymptotically Flat Spacetimes and the Impact of Mass-Dependent Variations}

\author{Vitalii Vertogradov}
\email{vdvertogradov@gmail.com}
\affiliation{Physics department, Herzen state Pedagogical University of Russia,
48 Moika Emb., Saint Petersburg 191186, Russia} 
\affiliation{SPB branch of SAO RAS, 65 Pulkovskoe Rd, Saint Petersburg
196140, Russia}

\author{Ali \"Ovg\"un
}
\email{ali.ovgun@emu.edu.tr}
\affiliation{Physics Department, Eastern Mediterranean
University, Famagusta, 99628 North Cyprus, via Mersin 10, Turkiye}

\begin{abstract}
Recent observations of black hole shadows have revolutionized our ability to probe gravity in extreme environments. This manuscript presents a novel analytic method to calculate, in leading-order terms, the key parameters of photon sphere and shadow radius. This method offers advantages for complex metrics where traditional approaches are cumbersome. We further explore the impact of black hole mass on the photon sphere radius, providing insights into black hole interactions with their surroundings. Our findings contribute significantly to black hole physics and gravity under extreme conditions. By leveraging future advancements in observations, such as the next-generation Event Horizon Telescope (ngEHT), this work paves the way for even more precise tests of gravity near black holes.
\end{abstract}

\date{\today}



\maketitle

\section{INTRODUCTION}
The groundbreaking observations of black hole shadows have ushered in a new era for probing gravity in extreme environments. In the 1970s, physicist James Bardeen introduced the idea of a black hole shadow, which has since become a major focus in astrophysics research \cite{bardeen1968non}. This shadow forms because light bends around the black hole, creating a dark area against the background of bright surrounding matter \cite{Luminet:1979nyg,Falcke:1999pj}. By studying this shadow, scientists can learn more about black holes, including their mass, spin, and the structure of spacetime nearby. Recent advances in observation, like very-long-baseline interferometry (VLBI) and the Event Horizon Telescope (EHT), have enabled direct imaging of the shadow of the supermassive black holes at the center of the M87  and and Sgr A* galaxies \cite{EventHorizonTelescope:2019dse,EventHorizonTelescope:2022wkp}. This has opened up new opportunities to test theories about gravity and the nature of black holes \cite{Vagnozzi:2022moj,Vertogradov:2024qpf,Perlick:2015vta,Virbhadra:2024pru,Virbhadra:2022iiy,Virbhadra:2007kw,Virbhadra:1999nm,Virbhadra:1998dy,Zakharov:2021gbg,Zakharov:2023lib,Abdujabbarov:2016hnw,Atamurotov:2013sca,Vagnozzi:2020quf,Claudel:2000yi,Hod:2012ax,Hod:2013mgr,Kuang:2022xjp,Konoplya:2016jvv,Johannsen:2011dh,Cardoso:2014rha,Konoplya:2019goy,Tsupko:2022kwi,Okyay:2021nnh,Lu:2019zxb,Tsukamoto:2014tja,Shaikh:2018lcc,Lambiase:2023zeo,Heydarzade:2023gmd,Allahyari:2019jqz}.

While General Relativity (GR) has successfully described many astronomical observations with high precision, there are compelling reasons to consider extending this framework. The theory has faced challenges in predicting phenomena such as the accelerating expansion of the Universe, the existence of dark matter, and its compatibility with quantum field theory. On the other hand, the direct observation of supermassive black holes offers a unique opportunity to test and constrain modifications to GR in their vicinity. Future projects like the Next Generation Event Horizon Telescope (ngEHT), which plans to incorporate satellites to increase baseline and image resolution, will provide even more precise measurements of supermassive black holes properties \cite{Tiede:2022grp,Ayzenberg:2023hfw}, such as spin. These observations will be crucial for testing alternative theories of gravity and improving our understanding of the fundamental forces governing the universe.

The photon sphere, also referred to as the circular photon orbit or light ring, and the shadow radius play crucial roles in the field of black hole physics \cite{Perlick:2021aok}. These quantities are central to the examination of particle trajectories, gravitational lensing effects \cite{Ishihara:2016vdc,Kudo:2022ewn}, and optical phenomena related to black holes. Following the EHT Collaboration's landmark observations of highly detailed black hole images, interest in studying photon spheres and black hole shadows has surged within the physics and astronomy communities. Traditionally, these values are determined through calculations involving the effective potential governing the motion of test particles within a black hole's spacetime. However, in certain scenarios, deriving analytic results can be exceedingly challenging, necessitating the use of numerical calculations. In this paper, we present a novel approach for determining the photon sphere and shadow radius in leading order terms.

On the other hand, comprehending the impact of mass on black hole characteristics, especially the radius of the photon sphere, holds paramount importance for advancing our understanding of black hole physics in realistic scenarios, particularly in the context of accretion processes around black holes where accretion refers to the process by which matter, such as gas, dust, or other material, is drawn in and accumulates around a massive object due to the object's gravitational pull \cite{Tan:2023ngk,Koga:2022dsu,Solanki:2022glc,Mishra:2019trb,Bambi:2015kza,Cunha:2018acu,Lima:2021las}. When it comes to black holes, accretion is a crucial phenomenon that occurs when matter from a surrounding disk or cloud falls toward the black hole. As this matter spirals inward, it forms an accretion disk—a structure of rapidly orbiting material that emits energy, often in the form of X-rays, as it is heated by friction and other processes. The mass of a black hole determines its gravitational pull, dictating where light is trapped in spacetime. Studying how the photon sphere radius changes with mass can reveal valuable insights into how black holes interact with their surroundings.

This paper has two primary aims. The first is to introduce a novel analytic method for calculating the photon sphere and shadow radius in leading-order terms. This method is particularly valuable for understanding the properties of metrics where standard methods are not applicable. The second aim is to investigate how the mass of a black hole influences the radius of its photon sphere. By studying the variation of the photon sphere radius with mass, we seek to comprehend how a black hole's gravitational field is impacted by its mass-dependent characteristics. This work has the potential to enhance our understanding of black hole physics and to contribute significantly to our knowledge of gravity in extreme environments.

This work is structured as follows: In Section II, we introduce a novel general approach for calculating the photon sphere and shadow radius. We then provide several illustrative examples. Section III outlines a method for estimating the change in the radius of a photon sphere with mass and applies this method to a selection of examples. Finally, we conclude with a summary in Section IV.

\section{GENERAL APPROACH FOR
ASYMPTOTICALLY FLAT METRIC}

We consider the asymptotically flat spherically-symmetric spacetime in the form
\begin{equation} \label{eq:metric1}
ds^2=-f(r)dt^2+f(r)^{-1}dr^2+r^2d\Omega^2,
\end{equation}
where $f(r)$ is a lapse function and $d\Omega^2=d\theta^2+\sin^2\theta d\varphi^2$ is the metric on the unit two-sphere.
We can extend the concept of expanding the photon sphere, where we express it as a Schwarzschild solution with small corrections, to the metric function. By considering a metric that approaches flat spacetime at infinity, we can expand it using negative powers of a small parameter 
\begin{equation} \label{eq:lapse}
f(r)=1-\frac{2 M}{r}+\sum_{i=2}^n\frac{\alpha_{i}}{r^{i}}+\mathcal{O}\left(\frac{1}{r^{n+1}}\right)    
\end{equation}

\textcolor{black}{Note that $M$ represents the mass of the black hole.} The parameters $\alpha_i$ represent the differences between a given metric and the Schwarzschild metric. The spacetime described by Equation \eqref{eq:metric1} is static and spherically symmetric. Along null geodesics, which describe the paths of light rays, there are two constants of motion: the energy per unit mass $E$ and the angular momentum per unit mass $L$. In the equatorial plane, where $\theta=\frac{\pi}{2}$, these constants of motion can be expressed as follows: 
\begin{eqnarray} \label{eq:energy}
E&=&f(r)\frac{dt}{d\lambda},\nonumber \\
L&=&r^2\frac{d\varphi}{d\lambda},
\end{eqnarray}
where $\lambda$ is an affine parameter. By using condition for null geodesics $g_{ik}\frac{dx^i}{d\lambda}\frac{dx^k}{d\lambda}=0$, one obtains the radial component $\frac{dr}{d\lambda}$ in the form
\begin{eqnarray} \label{eq:potential}
\left(\frac{dr}{d\lambda}\right)^2+V_{eff}(r)=0,\nonumber \\
V_{eff}=f(r)\frac{L^2}{r^2}-E^2,
\end{eqnarray}
\textcolor{black}{where $E$ and $L$ represent the energy per unit mass and angular momentum per unit mass, respectively and $V_{eff}(r)$ is an effective potential.} \textcolor{black}{The effective potential \( V_{\text{eff}} \) plays a crucial role in determining the behavior of photon orbits around a black hole. It encapsulates the combined effects of the black hole's gravitational field and the centrifugal barrier experienced by the photons. By analyzing \( V_{\text{eff}} \), we can identify the radii at which photons can have circular orbits (the photon sphere) and understand the stability of these orbits. The maxima and minima of \( V_{\text{eff}} \) correspond to the potential barriers and wells that photons encounter, thus shaping their possible trajectories and influencing the observed black hole shadow. }
The radius of a photon sphere $r_{ph}$ should satisfy the following two conditions
\begin{equation} \label{eq:conditions}
V_{eff}(r_{ph})=0,~~ V'_{eff}(r_{ph})=0.
\end{equation}
The second condition \eqref{eq:conditions} give for the lapse function \eqref{eq:lapse}
\begin{equation} \label{eq:sec_condition}
2r^n-6Mr^{n-1}+\sum_{i=2}^n (i+2)\alpha_i r^{n-i}=0.
\end{equation}
This algebraic equation has $n$ roots. However, we are interested only in maximal real root of the equation \eqref{eq:sec_condition} which has small \textcolor{black}{deviation} from $r_1=3M$ and we will consider only this case\footnote{We assume, that $\alpha_i\ll 1$.}.
\\ \\
\fbox{\begin{minipage}{25em}\textbf{ Theorem (Photon Sphere):} In order to estimate the influence $\alpha_i$ on the radius of a photon sphere, we assume the radius of a photon sphere $r_{ph}$ to be in the form
\begin{equation} \label{eq:radius1}
r_{ph}=r_{1}+\sum_{i=2}^n \alpha_ir_i+\mathcal{O}\left(\alpha_{n+1} \right)
\end{equation}

substituting it into the defining equation \eqref{eq:sec_condition} and expanding it up to order $\mathcal{O}(\alpha_i^2)$, we obtain for $r_i$ the relation:

\begin{equation}
r_i=-\frac{i+2}{6(3M)^{i-2}}
\end{equation}

Therefore, whether the radius increases or decreases depends on the sign of \( \alpha_i \): a positive sign leads to a decrease (as in the Reissner-Nordstrom case), while a negative sign corresponds to an increase.
\end{minipage}}
\\ \\

\textcolor{black}{The photon sphere theorem's derivation involves solving the equation for the effective potential's extrema, which determines the stable circular photon orbits. These solutions lead to the expression for the photon sphere radius in terms of the black hole's mass, charge and other parameters, encapsulated in the coefficients \(r_i\). These coefficients are critical as they directly relate to the properties of the black hole.}
\\
Now, we estimate how the radius of a shadow changes. We can write
\begin{equation} \label{eq:shadow}
b_{cr}^2=\frac{r_{ph}^2}{f\left(r_{ph}\right)}.
\end{equation}
By using the fact that at $r=r_{ph}$ the condition $f'r-2f=0$ is held, we can rewrite \eqref{eq:shadow} as
\begin{equation} \label{eq:shadow2}
b_{cr}^2=\frac{2r_{ph}}{f'\left(r_{ph}\right)}.
\end{equation}

Assuming lapse function $f(r)$ in the form \eqref{eq:lapse}, using the radius $r_{ph}$ in the form \eqref{eq:radius1}, neglecting terms of order $\mathcal(O)\left(\alpha_I^2\right)$, we arrive, after some algebraic calculations at
\\ \\
\fbox{\begin{minipage}{25em}\textbf{ Theorem (Shadow Radius):} 
\begin{equation} \label{eq:shadow_final}
b_{cr}^2=27M^2+27M\sum_{i=2}^n\alpha_ir_i-\frac{\left(27M^2+27M\sum_{i=2}^n \alpha_i r_i\right) \xi}{2Mr_{ph}^{n-1}-\xi},
\end{equation}
where
\begin{equation}
\xi=\sum_{i=2}^n i\alpha_ir_{ph}^{n-i}.
\end{equation}
\end{minipage}}
\\ \\
The equation \eqref{eq:shadow_final} defines the radius of a black hole shadow for an arbitrary asymptotically flat spacetime. The visible angular size of a shadow which can be seen by an observer far from black hole, i.e. $r_o\gg M$ ($r_o$ is a distance between an observer and black hole.) is given by (for asymptotically flat spacetimes  $f(r_o) \to 1$ \cite{Vagnozzi:2022moj}

\begin{equation}
\sin ^{2} \omega_{sh}=b_{cr}^{2}\frac{f(r_o)}{r_o}
\end{equation}

\begin{equation}
 R_{sh} \approx  b_{cr}. 
\end{equation}

B employing the first-order WKB formula and expanding the values for $\omega$ as a series of $\alpha_i$, it is also possible to determine the quasinormal modes in the eikonal regime \cite{Churilova:2019jqx}:
\begin{equation}
\begin{array}{c}\omega=\frac{\left(\ell+\frac{1}{2}\right)}{3 \sqrt{3} M}\left(1+\frac{\alpha_{2}}{6 M^{2}}+\frac{\alpha_{3}}{18 M^{3}}+\frac{\alpha_{4}}{54 M^{4}}+\frac{\alpha_{5}}{162 M^{5}}\right) \\ -i \frac{\left(n+\frac{1}{2}\right)}{3 \sqrt{3} M}\left(1+\frac{\alpha_{2}}{18 M^{2}}-\frac{\alpha_{3}}{27 M^{3}}-\frac{\alpha_{4}}{27 M^{4}}-\frac{11 \alpha_{5}}{486 M^{5}}\right) \\ \quad+\mathcal{O}\left(\frac{1}{\ell+\frac{1}{2}}\right).\end{array}
\end{equation}

\textcolor{black}{As a summary, our approach begins with identifying the general approach for asymptotically flat black hole spacetime metric and deriving the geodesic equations for photon motion. We then analyze the effective potential for radial motion to find the conditions for circular photon orbits, defining the photon sphere radius. Subsequently, we compute the apparent shadow radius as seen by a distant observer, using the impact parameter for light rays that approach the photon sphere. Finally, we examine how the photon sphere and shadow radii depend on various black hole parameters, such as mass and charge.}

\subsection{Example: Reissner-Nordstrom Black Hole}

We first give the example of Reissner-Nordstrom black hole. The metric function for
Reissner-Nordstrom black hole solutions is:

\begin{eqnarray}
  f (r) = 1 - \frac{2 M}{r} + \frac{Q^2}{r^2} .  \label{eq:metricRN}
\end{eqnarray}

where $Q$ is the black hole's  charge. Using the Eq.\ref{eq:lapse}, we can obtain the $\alpha$ coefficient for for magnetically charged black hole
as
\begin{eqnarray}
  \alpha_2 = Q^2.
\end{eqnarray}

In order to estimate the influence $\alpha_i$ on the radius of a photon
sphere, we assume the radius of a photon sphere $r_{ph}$ using the general approach: 
\begin{equation}
  \label{eq:radiusRN} r_{ph} =3 M-\frac{2 Q^2}{3}.
\end{equation}

Note that analytic result of photon sphere of Reissner-Nordstrom black hole is 

\begin{equation}
  \label{eq:radiusRNanalytic} r_{ph} =\frac{1}{2} \left(\sqrt{9 M^2-8 Q^2}+3 M\right).
\end{equation}

\begin{figure}
    \centering
\includegraphics[scale=0.7]{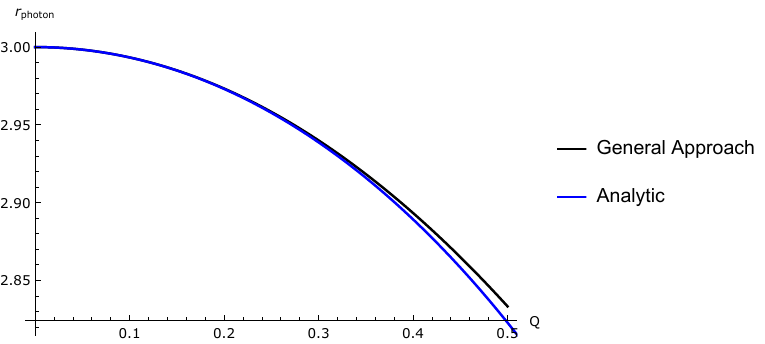}
    \caption{The Figure shows the comparison between general approach and analytic method for the photon sphere of Reissner-Nordstrom black hole.}
    \label{fig1}
\end{figure}

In the Fig. \ref{fig1}, it is seen how change of $Q $ affect the photon sphere. It is evident that for small values of $Q$, the two approaches align closely.

Now, we estimate how the radius of a shadow changes. The shadow radius for the Reissner-Nordstrom black hole is calculated using the general approach as:
\begin{equation} \label{eq:shadowapproach}
R_{sh}=-\frac{2 Q^2 \left(27 M^2-18 M Q^2\right)}{2 M \left(3 M-\frac{2 Q^2}{3}\right)-2 Q^2}+27 M^2-18 M Q^2.
\end{equation}

Note that analytic result of shadow radius for the Reissner-Nordstrom black hole is 

\begin{equation}
  \label{eq:shadowRNanalytic} R_{sh}=\frac{1}{2} \sqrt{-\frac{\left( \Lambda \right)^4}{4 M \left( \Lambda \right)+4 Q^2-6 \left( \Lambda \right)}}.
\end{equation}
where $\Lambda=\sqrt{9-8 Q^2}+3.$

\begin{figure}
    \centering
\includegraphics[scale=0.7]{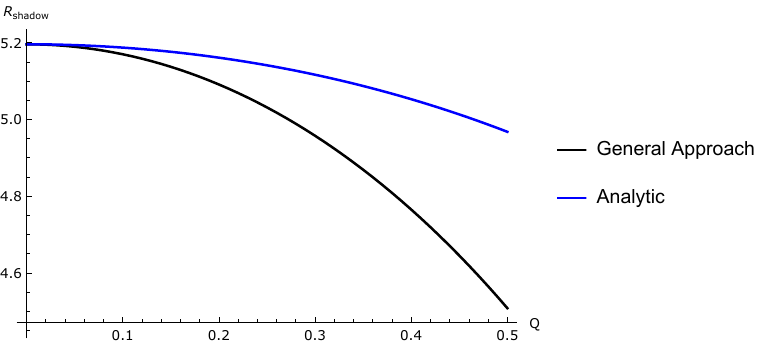}
    \caption{The Figure shows the comparison between general approach and analytic method for the shadow radius of Reissner-Nordstrom black hole.}
    \label{fig2}
\end{figure}

In the Fig. \ref{fig2}, it is seen how change of $Q $ affect the shadow radius of Reissner-Nordstrom black hole using two different method. It is observed that for small values of $Q$, the two approaches are in agreement. \textcolor{black}{Hence, for the Reissner-Nordstrom black hole, which includes electric charge in addition to mass, the results highlight how the presence of charge affects the photon orbits and shadow compared to the uncharged Schwarzschild case.}

\subsection{Example: Bardeen Black Hole}

Bardeen introduced an exact black hole solution in 1968 \cite{bardeen1968non,Vagnozzi:2022moj} that does not exhibit singularities at its origin, meaning that the curvature scalar does not tend to infinity. The line element describing the Bardeen regular black hole can be expressed as follows:

\begin{eqnarray}
  f (r) = 1-\frac{2 M r^2}{\left(e^2+r^2\right)^{3/2}}, 
\end{eqnarray}
Here, $M$ represents the mass, and $e$ denotes the monopole charge of a self-gravitating magnetic field sourced by non-linear electrodynamics.

The function $f(r)$ can be approximated asymptotically as
\begin{equation}
f(r) \approx 1-\frac{2 M}{r}+\frac{3 e^2 M}{r^3}-\frac{15 e^4 M}{4 r^5}+O\left(e^5\right)
\end{equation}

Using the Eq.\ref{eq:lapse},
we can obtain the $\alpha$ coefficient for Bardeen black hole
as
\begin{eqnarray}
  \alpha_3 =3 e^2 M, 
    \alpha_5 =-\frac{15 e^4 M}{4}, 
\end{eqnarray}
In order to estimate the influence $\alpha_i$ on the radius of a photon
sphere, we assume the radius of a photon sphere $r_{ph}$ 
\begin{equation}
  \label{eq:radius1new3} r_{ph} =3 M+\frac{35 e^4}{216 M^2}-\frac{5 e^2}{6}+O\left(e^5\right). 
\end{equation}
In the Fig. it is seen how change of $e $ affect the photon sphere.

Now, we estimate how the radius of a shadow changes. The shadow radius for the Bardeen black hole is calculated using the general approach as:
\begin{eqnarray} \label{eq:shadowapproachex3}
R_{sh}=3 \sqrt{3} M-\frac{\sqrt{3} e^2 (5 M+3)}{4 M}+O\left(e^5\right).
\end{eqnarray}

The shadow radius of the Bardeen black hole decreases as the monopole charge \( e \) increases.

\subsection{Example: Magnetically charged Einstein-Euler-Heisenberg Black Hole}

Another example is the magnetically charged black hole solutions derived from the Einstein-Euler-Heisenberg nonlinear electrodynamics, which represents a low-energy limit of Born-Infeld electrodynamics. The metric function for magnetically charged black hole solutions within this theory is given by \cite{Allahyari:2019jqz}:

\begin{eqnarray}
  f (r) = 1 - \frac{2 M}{r} + \frac{q_m^2}{r^2} - \frac{2 \mu}{5} 
  \frac{q_m^4}{r^6} \hspace{0.17em},  \label{eq:metricmagneticallyEEH}
\end{eqnarray}

where $q_m$ signifies the black hole's magnetic charge, a defining characteristic, while $\mu$ represents the coupling parameter in the Einstein-Euler-Heisenberg nonlinear electrodynamics (NLED), as outlined in Equation (\ref{eq:metricmagneticallyEEH}). Notably, for $\mu \neq 0$, magnetic charges exceeding $q_m > 1$ are allowable.

Utilizing Equation (\ref{eq:lapse}), we can derive the $\alpha$ coefficient for magnetically charged black holes as:
\begin{eqnarray}
  \alpha_2 = q_m^2, \\
  \alpha_6 = \frac{- 2 \mu}{5} q_m^4 . 
\end{eqnarray}
In order to estimate the influence $\alpha_i$ on the radius of a photon
sphere, we assume the radius of a photon sphere $r_{ph}$ 
\begin{equation}
  \label{eq:radius1new} r_{ph} = 3 M-\frac{8 \mu  q_m^4}{1215 M^4}-\frac{2 q_m^2}{3}. 
\end{equation}

Now, we estimate how the radius of a shadow changes. The shadow radius for magnetically charged black hole is calculated using the general approach as:
\begin{eqnarray} \label{eq:shadowapproachex2}
R_{sh}=\frac{2 \sqrt{\frac{2}{5}} M^{3/2} \sqrt{-\frac{\mu  q_m^4}{M^3}}}{3 M^{3/2}}-\frac{81 M^{3/2} \left(\sqrt{\frac{5}{2}} q_m^2\right)}{4 \left(M^{3/2} \sqrt{-\frac{\mu  q_m^4}{M^3}}\right)} \notag \\+O\left(M^{5/2}\right).
\end{eqnarray}

The photon sphere and shadow radius of the magnetically charged black hole decreases as the magnetic charge \( q_m \) increases.

\section{ESTIMATION HOW RADIUS OF A PHOTON SPHERE CHANGES WITH MASS}
In this section, we find a method to estimate how radius of a photon sphere changes with mass. For this purpose we consider general static spherically symmetric spacetime in Eddington-Finkelstein coordinates
\begin{eqnarray} \label{eq:spacetime}
ds^2&=&-f(r)dv^2+2dvdr+r^2d\Omega^2,\nonumber \\
f(r)&=&1-\frac{2M(r)}{r}.
\end{eqnarray}
This spacetime is supported with anisotropic energy-momentum tensor of the  form 
\begin{eqnarray} \label{emt}
\rho=-T^0_0=T^1_1=\frac{2M'}{r^2},\nonumber \\
P=T^2_2=T^3_3=-\frac{M''}{r}.
\end{eqnarray}
The radial component of the four-vector tangent to null geodesic is given by
\begin{eqnarray} \label{eq:radial}
\left(\frac{dr}{d\lambda}\right)^2+V_{eff}=0,\nonumber \\
V_{eff}=E^2-\left(1-\frac{2M}{r}\right)\frac{L^2}{r^2}.
\end{eqnarray}
Where $V_{eff}$ is an effective potential and $E$ and $L$ are energy and angular momentum respectively.
To find unstable photon orbit, one needs to satisfy two conditions \eqref{eq:conditions}. The second condition gives us a equation
\begin{equation} \label{eq:prom1}
3M-M'r-r=0.
\end{equation}
Now, we will find out how a radius of photon sphere $r_{ph}$ changes if we consider small change of mass of a black hole. For this purpose, we introduce small dimensionless parameter $|\alpha| \ll 1$ and the mass of black hole becomes $M+\alpha M$. Where $\alpha >0$ corresponds to accretion and $\alpha <0$ to radiation processes respectively.
We write the effective potential in the form
\begin{equation}
E^2-\left(1-\frac{2M}{r}\right)\frac{L^2}{r^2}\left(1-\frac{2\alpha M}{r-2M}\right).
\end{equation}
For simplicity, we introduce two new functions
\begin{eqnarray} \label{eq:def}
F(r)&\equiv &-\frac{r-2M}{r^3},\nonumber \\
G(r)&\equiv &\frac{2M}{r-2M}.
\end{eqnarray}
Then the second condition \eqref{eq:conditions} gives
\begin{equation} \label{eq:first}
F'(1-\alpha G)-\alpha G'F=0.
\end{equation}
Now we assume that the radius of a photon sphere is given by
\begin{equation} \label{eq:second}
r_{ph}=r_0+\alpha r_1,
\end{equation}
where $r_0$ is the radius of a photon sphere before accretion or radiation processes. Substituting \eqref{eq:second} into \eqref{eq:first} and expanding it with respect to alpha we find
\begin{equation} \label{eq:third}
r_1=\frac{F(r_0)G'(r_0)}{F''(r_0)}.
\end{equation}
Substituting \eqref{eq:def} into \eqref{eq:third} and evaluating it on the photon sphere, we find
\begin{equation} \label{eq:prom2}
r_1=\frac{r_0}{1-2M'(r_0)+M''(r_0)r_0}.
\end{equation}
The dominant energy condition demands $\rho \geq |P|$. From \eqref{emt}, we can  write $2M'\geq |-M''r|$ which gives us the dominator of \eqref{eq:prom2} as $1-4M'$. From \eqref{eq:prom1}, we know that $M'=\frac{3M}{r_0}-1$. Thus we have in the denominator of \eqref{eq:prom2} $5-\frac{12M(r_0)}{r_0}$. 

In the paper~\cite{Hod:2020pim}, it was proved that the radius of a photon sphere should be at least $\frac{3}{2}r_h$, where $r_h$ is the event horizon location. If we substitute it and use the fact that $\frac{2M}{r_h}\leq 1$ then we find that denominator of \eqref{eq:prom2} less than 1. This fact allows us to state that $r_1$ is always positive. 
\\ \\
\fbox{\begin{minipage}{23em}\textbf{ Theorem (Photon Sphere and Mass):}
This implies a direct relationship between the mass and the radius of the photon sphere: as the mass increases, the radius of the photon sphere also increases; conversely, if the mass decreases, so does the radius of the photon sphere $r_{ph}$. \end{minipage}}
\\ \\ 

We can show in the same manner that even we consider the equation of state $P=w(r)\rho$ then this condition is held for an arbitrary $w(r)$.
The mass function $M(r)$ depends upon several parameters which describe a black hole. 
\\ \\ 

 \textbf{For example in Reissner-Nordstrom case} it depends on both constant mass $m$ and electric charge $Q$. It may happen, that by increases the mass of black hole and electric charge the radius of photon sphere can decrease. As an example let's consider the Reissner-Nordstrom spacetime in which mass function is given by
 
\begin{equation} \label{eq:rn}
M(r,m,Q)=m-\frac{Q^2}{2r}.
\end{equation}

And we consider small positive change in mass and electrical charge by $m=m+\alpha m$ and $Q+\beta Q$, where $\alpha \ll 1$ and $\beta \ll 1$ are small positive constants. By substituting these changes into the mass function \eqref{eq:rn} and expanding it with respect to $\alpha$ and $\beta$, we arrive at \\
\begin{eqnarray} \label{eq:mass}
M(r,m+\alpha m, Q+\beta Q) \notag \\\approx m+\alpha \frac{\partial M(r, m, Q)}{\partial m} +\beta \frac{\partial M(r,m,Q)}{\partial Q}.
\end{eqnarray}

By choosing parameters $\alpha, \beta, m$ and $Q$ we can  make two last terms be negative. It means that we increases mass parameter $m$ but decreases the mass function $M(r)$ which means that the radius of a photon sphere and shadow decrease.

Now, we shall prove that it is possible only when the weak energy condition is violated.
For this purpose, we should understand the nature of parameters $\alpha$ and $\beta$.
We consider the accretion  or radiating processes. It means that the spacetime should be dynamical one. We write it in the form
\begin{equation}
ds^2=-\left(1-\frac{2M(v,r)}{r}\right)dv^2+2\varepsilon dvdr+r^2d\Omega^2.
\end{equation}
Where $M(v,r)$ is the mass function of both time $v$ and radial coordinate $r$ and $\varepsilon =\pm 1$ depending on ingoing or outgoing flux respectively.
The mass function change is described by $T^1_0$ component of energy-momentum tensor, which reads
\begin{equation}
T^1_0=2\varepsilon \frac{\dot{M}}{r^2}.
\end{equation}
In order to satisfy the weak energy condition this flux must be positive which means $\dot{M}\geq 0$ for accretion $\varepsilon=+1$ and negative $\dot{M}\leq 0$ for radiation $\varepsilon=-1$. However, the mass function depends on parameters $m$ and $Q$ which should vary  in time, i.e. they should be the functions of time. We consider the accretion process, i.e. $\varepsilon=+1$. It means
\begin{eqnarray}
\dot{M}=\frac{\partial M(r, M(v), Q(v))}{\partial m}\frac{dM(v)}{dv} \notag\\+\frac{\partial M(r, M(v), Q(v))}{\partial Q}\frac{dQ(v)}{dv}.
\end{eqnarray}
Comparing this expression with \label{eq:mass}, we find
\begin{eqnarray}
\alpha &=& \frac{dM(v)}{dv},\nonumber \\
\beta&=& \frac{dQ(v)}{dv}.
\end{eqnarray}

If the mass function $M(r)$ decreases it means that $\dot{M}$ decreases which violate weak energy condition.
We prove it only for two parameters but this can be extended for an arbitrary amount of parameters. 
We can state it as a theorem.
\\ \\
\fbox{\begin{minipage}{23em}\textbf{ Theorem:}
If the mass of a black hole increases while the weak energy condition holds, the size of the black hole's shadow also increases. \end{minipage}}

\subsection{Example: Hairy Schwarzschild Black Hole}
Our first example is the hairy Schwarzschild black hole, obtained through gravitational decoupling as described in the paper by Ovalle et al.~\cite{Ovalle:2020kpd}.
\textcolor{black}{ The gravitational decoupling refers to a scenario where the gravitational effects of a black hole can be decoupled from the effects of other fields or matter surrounding the black hole. Black hole mass parameter can be influenced by additional fields or interactions, leading to a more complex mass function than in the standard Schwarzschild solution.} It has the lapse function and
the mass function has the following form
\begin{eqnarray}
  f (r) = 1 - \frac{2 M (r)}{r}, 
\end{eqnarray}
and
\begin{eqnarray} \label{eq:mass_hairy}
M(r)&=&m+\frac{\bar{\alpha} l}{2}-\frac{\bar{\alpha}  r}{2}e^{-\frac{r}{M}}.
\end{eqnarray}

Here, $m=const.$ is the mass of a black hole, $\bar{\alpha}>0$ is coupling
constant and $l>0$ is related to a primary hair. We consider changes in
mass and primary hair as
\begin{eqnarray} \label{changes}
m\rightarrow m+\varepsilon m, ~~ \varepsilon \ll 1,~~
\varepsilon>0,\nonumber \\
l \rightarrow l+\beta l,~~ \beta \ll 1,~~ \beta > 0.
\end{eqnarray}

We can write the mass function up to order $\mathcal{O}\left(
\varepsilon^2\right)$, $M(r) \to  \mathcal{M}(r)$ as


\begin{eqnarray}
 \mathcal{M}(r)=  \frac{\bar{\alpha}   \beta  l}{2}+\frac{\bar{\alpha}  l}{2}-\frac{1}{2} \bar{\alpha}   r e^{-\frac{r}{m}}+m\notag \\+\epsilon  \left(m-\frac{\bar{\alpha}   r^2 e^{-\frac{r}{m}}}{2  m}\right)+O\left(\epsilon ^2\right), \end{eqnarray}

 and

 \begin{eqnarray}
 \mathcal{M}(r)<  M(r)+\frac{\bar{\alpha}   \beta  l}{2}+\epsilon  \left(m-\frac{\bar{\alpha}   r^2 e^{-\frac{r}{m}}}{2  m}\right)+O\left(\epsilon ^2\right). \end{eqnarray}
 
As one can see, under some conditions, the new mass function
$\mathcal{M}(r)$ might be less than $M(r)$, i.e. the radius of a photon
sphere decreases despite the mass of a black hole grows. However, it
might happen only if $\bar{\alpha}  >1$.

\subsection{Example: Kiselev Black Hole}

Kiselev black hole~\cite{Kiselev:2002dx} is explicit example that the
radius of a photon sphere might decrease while the mass of a black hole
grows. The Kiselev solution has the lapse and mass functions of the form

\begin{eqnarray}
  f (r) = 1 - \frac{2 M (r)}{r}, 
\end{eqnarray}
and
\begin{eqnarray} \label{eq:kiselev}
M(r)&=&m+\frac{N}{2r^{3\omega}}.
\end{eqnarray}
Here $m=const.$ is a black hole mass, $N$ is a cosmological parameter
and \textcolor{black}{$\omega$ is a parameter of equation of state and we consider only
positive values of $\omega$. The parameter $\omega$ is a parameter of barotropic equation of state $P=\omega \rho$, where $P$ and $\rho$ are pressure and energy density of surrounded fluid respectively.} In this case, the cosmological parameter
$N$ should be always negative. We consider the positive changes of mass
and cosmological parameters
\begin{eqnarray} \label{changes2}
m\rightarrow m+\varepsilon m,~~ \varepsilon \ll 1,~~
\varepsilon>0,\nonumber \\
N\rightarrow N+\beta N,~~ \beta \ll1,~~ \beta>0.
\end{eqnarray}
In this case $M(r) \to  \mathcal{M}(r)$ as

\begin{equation}
 \mathcal{M}(r)= m+m \epsilon +\frac{1}{2} \beta  N r^{-3 \omega }+\frac{1}{2} N r^{-3 \omega }
\end{equation}
and one can write it
\begin{equation}
 \mathcal{M}(r)<M(r)+m \epsilon +\frac{1}{2} \beta  N r^{-3 \omega }.
\end{equation}

With negative $N$ one can always find such $\varepsilon$ and $\beta$
that $\mathcal{M}(r)<M(r)$.

\subsection{Example: Magnetically charged Einstein-Euler-Heisenberg Black Hole}

Another example is the magnetically charged black hole solutions within this theory, as described in \cite{Allahyari:2019jqz}
\begin{eqnarray}
  f (r) = 1 - \frac{2 m}{r} + \frac{q_m^2}{r^2} - \frac{2 \mu}{5} 
  \frac{q_m^4}{r^6} \hspace{0.17em},  \label{eq:metricmagneticallyEEH2}
\end{eqnarray}
where $q_m$ denotes the magnetic charge of the black hole, representing a distinct attribute, while $\mu$ signifies the coupling parameter in the Einstein-Euler-Heisenberg nonlinear electrodynamics (NLED), as detailed in Eq. (\ref{eq:metricmagneticallyEEH2}). It's noteworthy that for $\mu \neq 0$, magnetic charges exceeding $q_m > 1$ are admissible.

The metric function for magnetically charged black hole solutions can also be expressed in the following form:

\begin{eqnarray}
  f (r) = 1 - \frac{2 M (r)}{r}, 
\end{eqnarray}
where the function $m (r)$, in this case, is given by:
\begin{equation}
  \label{eq:example_4} M (r) = \frac{10 mr^5 + 2 \mu q_m^4 - 5 q_m^2 r^4}{10
  r^5} .
\end{equation}

On the other hand we can also estimate how radius of a photon sphere changes with mass

Here $m=const.$ is a black hole mass. and $q_m$ is the magnetic charge which is always positive. We consider the positive changes of mass
and charge parameters
\begin{eqnarray} \label{changes2}
m\rightarrow m+\varepsilon m,~~ \varepsilon \ll 1,~~
\varepsilon>0,\nonumber \\
q_m\rightarrow q_m+\beta q_m,~~ \beta \ll1,~~ \beta>0.
\end{eqnarray}

In this case $M(r) \to  \mathcal{M}(r)$ as

\begin{eqnarray}
\mathcal{M}(r)= \left(m \epsilon +m+\frac{\mu  q_m^4}{5 r^5}-\frac{q_m^2}{2 r}\right) \\+\beta  \left(\frac{4 \mu  q_m^4}{5 r^5}-\frac{q_m^2}{r}\right) \notag
\end{eqnarray}
which can be written as
\begin{eqnarray}
\mathcal{M}(r)< M(r) + m \epsilon +\beta  \left(\frac{4 \mu  q_m^4}{5 r^5}-\frac{q_m^2}{r}\right). 
\end{eqnarray}
With negative $\mu$ one can always find such $\varepsilon$ and $\beta$
that $\mathcal{M}(r)<M(r)$.

\subsection{Example:  Black holes in MOdified Gravity (scalar-tensor-vector gravity)}

Last example is the black holes in Modified Gravity (MOG black hole) proposed by Moffat in Ref.~\cite{Moffat:2005si}, commonly known as scalar-tensor-vector gravity. In this scenario, the metric function describing the resulting black hole solution is given by \cite{Moffat:2014aja}.

\begin{eqnarray}
  f (r) = 1 - \frac{2 M (r)}{r}, 
\end{eqnarray}
where the function $M (r)$, in this case, is given by:
\begin{equation}
  \label{eq:metricmodifiedgravity} M (r) = -\frac{\alpha ^2 m^2}{2 r}-\frac{\alpha  m^2}{2 r}+\alpha  m+m.
\end{equation}

Here, $\alpha$ is a parameter governing the strength of the effective gravitational coupling $G=G_N(1+\alpha)$, representing a universal hair.

Here $m=const.$ is a black hole mass. and $\alpha$ is a constant. We consider the positive changes of mass
and charge parameters
\begin{eqnarray} \label{changes2}
m\rightarrow m+\varepsilon m,~~ \varepsilon \ll 1,~~
\varepsilon>0,\nonumber \\
\alpha \rightarrow \alpha+\beta \alpha ,~~ \beta \ll1,~~ \beta>0.
\end{eqnarray}

In this case $M(r) \to  \mathcal{M}(r)$ as

\begin{eqnarray}
 \mathcal{M}(r)= -\frac{\alpha  \beta  m^2}{2 r}-\frac{\alpha  m^2}{2 r}+\alpha  \beta  m+\alpha  m+m\notag \\+\epsilon  \left(-\frac{\alpha  \beta  m^2}{r}-\frac{\alpha  m^2}{r}+\alpha  \beta  m+\alpha  m+m\right)
\end{eqnarray}
then we rewrite it

\begin{eqnarray}
 \mathcal{M}(r)<M(r) -\frac{\alpha  \beta  m^2}{2 r}+\alpha  \beta  m \notag \\+\epsilon  \left(-\frac{\alpha  \beta  m^2}{r}-\frac{\alpha  m^2}{r}+\alpha  \beta  m+\alpha  m+m\right)
\end{eqnarray}
With negative $\alpha$ one can always find such $\varepsilon$ and $\beta$
that $\mathcal{M}(r)<M(r)$.

\section{CONCLUSIONS}
 A shadow is a critical aspect of a black hole, providing valuable insights into its structure. Modern imaging techniques have enabled us to observe these shadows, prompting a deeper investigation into their characteristics, particularly regarding the influence of various black hole properties like accretion, radiation, and other parameters. Understanding the shadow is pivotal for validating or refuting black hole models. Furthermore, shadows are not exclusive to black holes; other compact objects like wormholes and naked singularities can also cast shadows, offering a means to probe these enigmatic entities~\cite{Heydarzade:2023gmd, Shaikh:2018lcc}. Shadows also serve as a tool to test the nature of the objects obscured behind them. However, due to the often complex spacetime structures, analytical calculations of photon spheres or shadows are frequently unfeasible, necessitating the use of numerical methods to study their properties.

In this study, we present a novel method for determining the radius of a photon sphere and shadow in any asymptotically flat spherically symmetric spacetime. Our method facilitates a straightforward comparison of photon sphere and shadow radii with those of a Schwarzschild black hole, providing valuable insights into the gravitational effects of various spacetime geometries. We validate our approach by applying it to several well-known examples, demonstrating its efficacy.

Moreover, we address a significant question regarding the shadow of a dynamical black hole. In scenarios involving accretion and radiation processes, the mass of the black hole and the surrounding spacetime geometry evolve dynamically. However, the absence of a timelike Killing vector in dynamical spacetimes poses challenges in reducing the second-order equations of motion to first-order ones. Consequently, numerical methods or the identification of additional symmetries become necessary. In this paper, we take initial steps towards analytically understanding shadows and photon spheres in dynamical spacetimes, focusing specifically on the accretion process.

\textcolor{black}{We show that variations in the mass function of the black hole correlate with changes in the radius of the photon sphere. However, it is crucial to distinguish between the mass function and the actual mass of the black hole, as the former depends on various parameters such as radius and mass. We illustrate this distinction with an example involving the accretion of charged particles, where despite an increase in mass, the radius of the photon sphere decreases due to a higher rate of electrical charge accumulation relative to mass. We rigorously prove, as a theorem, that such scenarios occur only when the weak energy condition is violated. This theorem lays the groundwork for the future development of analytical methods for calculating shadows cast by dynamical black holes, promising advancements in our understanding of these complex phenomena. Moreover, the potential observational implications, such as the deviations in shadow characteristics detectable by future astronomical observations.}

\textcolor{black}{Looking towards the future, our work lays the groundwork for further advancements in the field. The next-generation Event Horizon Telescope (ngEHT) promises even more precise tests of gravity near black holes. By continuously leveraging advancements in observational technology, we can refine our understanding of black holes, ultimately leading to a deeper grasp of the fundamental forces governing the universe.}

\acknowledgements
V. Vertogradov thanks the Basis
Foundation (grant number 23-1-3-33-1) for the financial support. A. {\"O}. would like to acknowledge the contribution of the COST Action CA21106 - COSMIC WISPers in the Dark Universe: Theory, astrophysics and experiments (CosmicWISPers) and the COST Action CA22113 - Fundamental challenges in theoretical physics (THEORY-CHALLENGES). We also thank TUBITAK and SCOAP3 for their support.

\bibliography{reference.bib}
	
\end{document}